# CENTRAL BANK DIGITAL CURRENCIES: THE ADVENT OF ITS IT GOVERNANCE IN THE FINANCIAL MARKETS.

**Carlos Alberto Durigan Junior** ; https://orcid.org/0000-0003-2185-493X
POLI USP

**Mauro De Mesquita Spinola** ; https://orcid.org/0000-0002-5147-9395
POLI USP

**Rodrigo Franco Gonçalves** ; https://orcid.org/0000-0003-2206-3136
UNIP

**Fernando José Barbin Laurindo** ; https://orcid.org/0000-0002-5924-3782
POLI USP

Title: *Central Bank Digital Currencies: The advent of its IT governance in the financial markets.*

**Abstract**

Central Bank Digital Currency (CBDC) can be defined as a virtual currency based on node network and digital encryption algorithm issued by a country which has a legal credit protection. CBDCs are supported by Distributed Ledger Technologies (DLTs), and they may allow a universal means of payments for the digital era. There are many ways to proceed, they all require central banks to develop technological expertise. Considering these points, it is important to understand the new IT governance in the financial markets due to CBDC and digital economy. Information Technology is an essential driver that will allow the new financial industry design. This paper has the objective to answer two questions through an updated Systematic Literature Review (SLR). The first question is What IT resources and tools have been considered or applied to set the governance of CBDC adoption? The second; Identify IT governance models in the financial market due to CBDC adoption. Bank for International Settlements (BIS) publications, Scopus and Web of Science were considered as sources of studies. After the strings and including criteria were applied, fourteen papers were analyzed. This paper finds many IT resources used in the CBDC adoption and some preliminary IT design related to the IT governance of CBDC, in the results and discussion section the findings are more detailed. Finally, limitations and future work are considered.

1. **Introduction**

According to the International Monetary Fund (IMF) Central Bank Digital Currency (CBDC) may be understood as a kind of legal tender in digital form, following the primary money functions (Kiff *et al*., 2020). CBDC is the evolution in money format from metal currency to metal-backed banknotes, and then to fiat money (Lee *et al*, 2021). An important difference between cash and electronic retail money is that the latter represents a claim on an intermediary, whereas the former is a direct claim on the central bank (BIS, 2020). A CBDC should allow central banks provide a universal means of payment for the digital era, safeguarding consumer privacy and preserving the private sector's primary role in retail payments and financial intermediation (BIS, 2020).

CBDC can foster competition among private sector intermediaries and serve as a basis for sound innovation in payments (Brunnermeier*, et al.* 2021). The advent of these new monies could reshape the nature of currency competition, the structure of the international monetary system, and the role of government-issued public money. In a digital economy, cash may effectively disappear, and payments may center around social and economic platforms rather than banks. Governments may need to offer central bank digital currency (CBDC) in order to retain monetary independence (Brunnermeier*, et al.* 2021).

Central banks can also improve the services they supply directly to users by keeping technological frontiers. CBDCs can serve both as a complementary means of payment that addresses specific use cases and as a catalyst for continued innovation in payments and commerce. For instance, in Sweden, where cash use has significantly declined, central bank proposed to local government that digital central bank money held by some people should also be given the status of legal tender (Sveriges Riksbank, 2019). In addition, Facebook has already started acting about its own crypto currency, the Libra. Thousands of fiat cryptocurrencies maintained on Blockchains by anonymous record-keepers have been launched lately (Brunnermeier, *et al.* 2021). Blockchain is a technology that allows a growing list of data structures (blocks) connected and secured by cryptography (Haber and Stornetta, 1990).

About technological aspects, a range of different operational arrangements could be conceivable. In some, the central bank hosts a database of retail balances (these might be anonymized), whereas in others it keeps track only of wholesale balances. There are many ways to proceed, but all require central banks to develop substantial technological expertise and some key terms about the design may consider interoperability among platforms, Blockchain and other Distributed Ledger Systems adoption (BIS, 2020).

It is still not clear if digital currencies will keep the traditional functions of paper money (store of value, medium of exchange, and unit of account). Digital currencies may compete exclusively as exchange media or exclusively as stores of value. Additionally, it is possible that transactions will be made between Digital Currency Areas (DCAs) linking the currency to usership of a particular digital network rather than to a specific country (Brunnermeier, *et al.* 2021).

Considering these, CBDCs have the potential to be the next step in the evolution of money, but a thoughtful approach is necessary. CBDC issuance is not so much a reaction to cryptocurrencies, it is rather a focused technological effort by central banks to pursue several public policy objectives. These objectives include financial inclusion, safety and integrity in digital payment, while encouraging continued innovation. CBDC can stablish a new payment mechanism that is interoperable by default, fosters competition and sets high standards for safety and risk management (Auer & Böhme, BIS 2021).

Considering the technological aspects, every form of digital money requires a distributed record-keeping system. This record-keeping system updates a shared state, which encodes how a given number of currency units is allocated to holders. Technical communication ensures that every component of the distributed system is up to date at least with the part of the shared state relevant to this component. In the case of CBDC, laws must be created to ensure that information encoded in this state is mapped to ownership of claims against the central bank. (BIS, 2020).

As stated in this introduction, it is possible to realize that information technology plays an important function in order to successfully establishes a CBDC adoption. Both IT resources and IT governance may influence in the design of the architectural aspects about the establishment of a CBDC, there is a close relationship between Information Technology and Digital Currencies. Considering the aforementioned aspects, this paper seeks to answer the following questions: 1) What IT resources and tools have been considered or applied in order to set the governance of CBDC adoption? 2) Identify IT governance models in the financial market due to CBDC adoption.

The two questions above consist the main objective of this paper. To answer them a Systematic Literature Review (SLR) is considered as a methodological approach. Following this Introduction this paper is structured in the following topics; Literature Review, Methodology, Results and Discussion, Conclusions, Limitations and Future Works.

## 2. Theoretical Background and Literature Review

Central banks have recently started considering issuing their own digital currency (CBDC). David Chaum set out his vision for anonymous electronic cash in 1985 (Chaum 1985). (Auer & Böhme, BIS 2021). It is also important the concerns about the operational aspects and the efficiency of the payment system. The side of retail payments, including onboarding for payment accounts, authorization, clearing, settlement, compliance with anti-money laundering (AML) and counter the financing of terrorism (CFT) rules are crucial operational issues. These tasks are arguably better handled by the private sector than the central bank (Borio (2019), Carstens (2019) and BIS (2020)). (Auer & Böhme, BIS 2021).

The economic design of a CBDC should allow commercial banks to keep their intermediation role between savers and investors. (Auer & Böhme, BIS 2021). A second consideration for a minimally invasive CBDC design regards the operational dimension and the efficiency of the payment system. As the customer-facing side of retail payments is better handled by the private sector than the central bank. (Auer & Böhme, BIS 2021). Technical architecture of a CBDC is defined by the role of the components of the distributed record-keeping system, their communication relations, and the question who is in control of each component. This design may differ in terms of the structure of legal claims and the record kept by the central bank. (Auer & Böhme, BIS 2021).

The CAP theorem in computer science (Gilbert and Lynch 2002) tells us that no distributed system can be available and consistent while parts of it are disconnected. Many existing electronic payment systems work around this technical impasse with an economic solution. They involve intermediaries, If the central bank would run the system itself, it would have to engage in this risk taking. (Auer & Böhme, BIS 2021). Technological resilience is achieved in the hybrid design via the central bank operating a backup infrastructure (hence the name hybrid – a payment system that can run on either a public or a private infrastructure). (Auer & Böhme, BIS 2021).

This is an area of ongoing technical research, primarily for the use in distributed ledgers, which are not necessarily the best choice to set up a CBDC infrastructure. It remains to be seen which new results in cryptography withstand the test of time and can be applied fruitfully in architectures suitable for CBDC (Auer & Böhme, BIS 2021). Central Banks can maintain record-keeping directly or outsource it to private sector and supervise it. To establish credible direct claims in electronic form, it is sufficient if the central bank has a view (ie read access) to an authoritative source of information. The record-keeping can be delegated to the private sector, who might either be allowed to use proprietary technology or be required to run some open protocols specified by a Regulator (Auer & Böhme, BIS 2021).

Many approaches proposed by the industry envision payment systems that feature intermediaries but seek to reduce dependence on them. For example, several CBDC prototypes are built on enterprise versions of distributed ledgers, such as Corda, Hyperledger, or Quorum (Auer *et al*, 2020). These versatile software packages were inspired by and borrow concepts from decentralized cryptocurrencies. Most central bank projects have good reasons for running them in configurations that resemble a redundant but centrally controlled database rather than Bitcoin. Some academic proposals are designed to break with conventional databases. They adapt selected principles of decentralized cryptocurrencies to the use case of CBDC (Auer & Böhme, BIS 2021).

Convertibility among monetary instruments and interoperability between platforms will be crucial in reducing barriers to trade and enabling competition. Digital currencies may also cause an upheaval of the international monetary system: countries that are socially or digitally integrated with their neighbors may face digital dollarization, and the prevalence of systemically important platforms could lead to the emergence of digital currency areas. The advent of digital currencies will have implications for the treatment of private money, data ownership regulation, and central bank independence. In a digital economy where most activity happen through networks with their own monetary instruments, a regime in which all money is convertible to CBDC would uphold the unit of account status of public money (Auer & Böhme, BIS 2021).

To understand what constitutes an independent currency it is important to define what it means for a payment instrument to belong to a currency. A collection of payment instruments forms an independent currency if the following two conditions are applied: (i) The payment instruments are denominated in the same unit of account. (ii) Each payment instrument within the currency is convertible into any other. (Auer & Böhme, BIS 2021).

The centrality of payments and data on social and commercial platforms may lead to an inversion of the current industrial organization of financial activities. Banks are the point of contact for all users of the payment system. In many countries, banks' dominance of financial activities extends even to the provision of insurance and asset management services. The financial system, and the way in which consumers store and exchange value, is organized around banks and credit. In a platform-based economy, this hierarchy could be overturned. Payments are at the center of any economic platform, and all other activities would organize themselves around the central payment functionality (Auer & Böhme, BIS 2021).

In this new type of financial hierarchy, traditional financial institutions such as banks could be replaced by fintech of payment systems. Fintech is a financial company mostly based on technological processes and resources. This type of industrial organization is already flourishing in some countries. In China an Alibaba's financial branch has become the world's largest money-market mutual fund. Sesame Credit, another subsidiary, has emerged as a dominant credit scoring system (Auer & Böhme, BIS 2021).

It is possible that in the future each digital currency will come bundled with many data services and be associated with a collection of economic activities occurring across borders. The traditional system of intermediation may be turned on its head, with payment providers sitting on top of subsidiaries that provide intermediation and other financial services (BIS 2021).

A central bank money that serves only as a medium of exchange is potentially vulnerable to technological change. Digitalization may allow to dispense from base money and settle payments differently. Inside large digital networks most transactions can be settled internally, thus bypassing central banks. According to Brunnermeier*, et al* (2021) the larger the network, the smaller its need for an outside settlement asset. Some projects launched by consortiums of banks, such as JP Morgan's JPM Coin or the Finality blockchain, could circumvent traditional settlement with reserves by building a network on which many types of payments, including cross-border payments, could be instantaneously finalized using tokens (Brunnermeier*, et al.* 2021).

CBDC may have the potential to replace physical cash in the future. CDBC may also complement electronic money and performs similar functions or act as a fully backed reserve. CBDC is a new type of electronic liabilities of the central bank, which can be used as a means of payment and value storage and retain most of the desirable characteristics. Due to the varying financial development stages and different needs of countries, there are many definitions from a variety of perspectives in literature (Kuo et. al, 2021).

In 2014, China started research on fiat digital currencies and achieved staged results in key parts such as theory and framework. The Bank of England proposed the world's first central bank digital currency in 2015, RSCoin (Danezis and Meiklejohn, 2015) Canada launched the Jasper project to study the Canadian dollar fiat digital currency in 2017 (Chapman *et al*, 2017). In the same year, the Monetary Authority of Singapore, European Central Bank, and Bank of Japan also carried out research on CBDC with the project named Ubin (Dalal *et al*, 2017) and STELLA (ECB and Bank of Japan, 2019). Tsai *et al*. proposed an account-based fiat digital currency in 2018 (Tsai *et al*, 2018), (Zhang *et al*, 2021).

In a blockchain network node architecture, the blockchain network architecture is divided into traditional peer-to-peer networks and structured modular networks. The former is used in most digital currencies such as Bitcoin (Zhang *et al*, 2021). About the traditional Blockchain architecture, digital currency systems tend to use a unified wallet software as a node carrier, which updated by a community or cooperative organization and keeping all nodes in fair conditions. Hence, a node usually keeps a complete ledger database and receives digital currency rewards from local mining. Besides, the node also stores the user's private key, initiates a transaction, acts as a proxy for others, and provides basic functions (consensus, encryption, decryption, hash operations, transaction pools, etc.). Therefore, all nodes in the architecture are comprehensive but difficult to make custom optimization (Zhang *et al*, 2021).

About the modular Blockchain Architecture, as the alliance chain itself is used in a semi-centralized condition, the nodes have more credibility between peers. The architecture design of alliance chain is more flexible, such as the Hyperledger Fabric (Androulaki *et al*, 2018). Especially, this architecture modularizes the functions of nodes and sets different permissions for them. When a fault occurs, the faulty part can be repaired separately in this way. If there is a demand, a type of node can be expanded alone to enhance the processing capacity. In the design, the repeated processing work is reduced, more resources are saved, thus the performance of the system is improved (Zhang *et al*, 2021).

Regarding the network architecture, the traditional digital currency architecture generally has the problems of insufficient operating efficiency and excessive resource consumption. Thus, it is not suitable for CBDC that requires high throughput. The high efficiency blockchain architecture represented by the alliance chain has excellent performance and modular architecture. But its application scenarios and some design concepts are not compatible with CBDC. Zhang *et al* (2021) proposed an architectural design based on the alliance chain. A customized structured and modular alliance chain architecture for CBDC was proposed. Different from using the unified wallet in traditional blockchain architecture, nodes are divided into types in our design. They are responsible for consensus, account processing, and UTXO processing. To improve the concurrent processing capability, a sliced storage solution was proposed by Zhang *et al* (2021) to optimize the data management, and the working scope of nodes with the same type are independent and concurrent. (Zhang *et al*, 2021).

IT Governance (ITG) is essential to an organization's success. As IT is associated with risk and value opportunities, ITG has become imperative for business organizations to meet the challenges presented by the business environment, IT stands for a competitive advantage. There are many studies about ITG (e.g., Calder & Moir, 2009; Calder, 2005; Willson & Pollard, 2009), and some reports from leading enterprises in ITG such as by ITGI and ISACA. However, these studies are not at the same level of ITG importance in some aspects such Critical Success Factors (CSFs) (Alreemy, *et al*, 2016).

COBIT 5 introduced CSFs for the ITG processes but they cannot be used as CSFs for the entire ITG implementation. ISO 38500 introduced six principles and it was the first inter-national standard explicitly addressed the governance of ICT. It is clear that the alignment between IT and business is an important aspect that should be considered in the implementation of IT (Alreemy, *et al*, 2016).

The Information Systems (IS) literature related to inter-firm IT governance deals primarily with the IS outsourcing environment Inter-firm IT governance usually includes contractual governance and relational governance (Cao, Mohan, Ramesh, & Sarkar, 2013). Both relational and contractual governance are necessary and effective governance mechanisms in the process of managing IT outsourcing (Deng, Mao, & Wang, 2013; Kim, Lee, Koo, & Nam, 2013; Tiwana, 2010) (Chi *et al*, 2017).

Lazic et al. (2011) found that IT governance is positively related to business performance through IT relatedness and business process relatedness. Prasad et al. (2012) suggest that IT governance structures contribute to firm performance through IT-related capabilities which improve the effectiveness and efficiency of the internal business processes. Yet, among these few works on the governance– performance link, there is no consensus as to exactly how IT governance enhances performance and it is still unclear by which precise mechanisms IT governance exerts its effects on firm performance (Ju Wu *et al*, 2015).

Prior research has identified and defined different kinds of IT capabilities (Bharadwaj, 2000; Bharadwaj et al., 1999; Bhatt & Grover, 2005; Feeny & Willcocks, 1998; Mithas, Ramasubbu, & Sambamurthy, 2011; Wade & Hulland, 2004), which is one family of organizational capabilities. More specifically, IT capabilities center on IT resources and practices, and have been shown to improve organizational performance (Bharadwaj, 2000; Chakravarty, Grewal, & Sambamurthy, 2013, *Apud* Turel *et al*, 2017).

To have a positive effect on an organization's competitive position and performance, an IT capability needs to be valuable, heterogeneously distributed, and imperfectly mobile (Mata, Fuerst, & Barney, 1995). Strategic alignment is defined as the degree to which the mission, objectives, and plans contained in the business strategy are shared and supported by the IT strategy (Reich & Benbasat, 1996). Moreover, organizational performance is one outcome of this strategic alignment (Henderson & Venkatraman, 1993; Sabherwal & Chan, 2001; Wu *et al*., 2015) (Turel *et al*, 2017).

IT Governance IT Governance is related to Corporate Governance and is concerned with the control and transparency of decisions in Information Technology, without disregarding mechanisms and processes to increase the effectiveness of IT (Peterson, 2004). The components of the Information Technology environments, such as networks, servers and applications, are increasingly complex, composed of more sophisticated items, and require a higher level of technological integration. The increasing integration of administrative and factory functions allows the creation of Enterprise Resource Planning (ERP) and Manufacturing Execution Systems (Amess *et al*., 2007).

The integration between suppliers and customers, with the creation of integrated supply chains, and the intensification of the relationship with customers, helps to increase the integration, sophistication and complexity of the IT environment (Fernandes; Abreu, 2008).In the view of Webb et al. (2006), the three elements most frequently associated with IT Governance in the literature are structures, control frameworks and processes. However, to define IT Governance, it involves a broader spectrum. The establishment of a "definitive definition" that observes such a "broad spectrum" and that does not limit or restrict the conceptual scope (Webb *et al*., 2006).

### 3. Methodological Approach

This paper considers a Systematic Literature Review (SLR). Considering the importance of Central Bank Digital Currencies (CBDC) in the context of the digital economy, it is important to understand the relationship between Information Technology (IT) governance and the dynamics of digital currencies. These new currencies also open the possibility of having decentralized finance (DEFI). Considering this, this paper seeks to answer two questions through a SLR methodological approach along with content analysis of the selected literature reviewed.

Questions to be answered:

1) What IT resources and tools have been considered or applied to set the governance of CBDC adoption?

2) Identify IT governance models in the financial market due to CBDC adoption.

Considering these questions this paper applies SLR approach (Tranfield, Denyer, & Smart, 2003) in combination with Kitchenham (2004) and Kitchenham *et al*. (2009). As suggested by these authors, the literature review can be subdivided into three main phases: planning the review, conducing the review and reporting it.

I. Planning the Review

The review considers the following meanings: "Central Bank Digital Currency" OR "CBDC" AND "Information Technology" AND "Governance". Sources considered: Scopus and Web of Science. These sources were considered because they are among the most used sources in academic environment. They do provide most of published literature. As the theme of this paper is intrinsically related to Central Banks, it was also searched (in addition) publications by the Bank for International Settlements (BIS) which is a Central Bank for Central Banks worldwide. Table 1 below details the criteria.

Table 1
**Including and Excluding Criteria**

| Including Criteria | Excluding Criteria |
|---|---|
| • Papers covering both Central Bank Digital Currency (CBDC) and Information technology.<br>• Academic papers and Books.<br>• Working Papers and Reports. | • Papers that do not cover relationship between CBDC and Information Technology or were not helpful to answer the two main questions.<br>• Papers about other fields (other than digital economy)<br>• Papers that were not fully (or free) available (whole document) in the sources searched. |

Source: Author.

II. Conducting the review

The search was performed using the Web of Science and Scopus scientific databases using the final strings in Table 2. Drawing on the methodological frameworks of Tranfield *et al*. (2003); Kitchenham (2004) and Kitchenham *et al*. (2009). For Scopus database the terms were searched in abstracts, titles, and keywords, without any other constraints. For Web of Science database, the strings were searched in "Topics". In this phase, the following articles information were exported: title, authors, abstract, publication year, keywords, source title, document type and language. Thus, papers exported metadata were saved on Microsoft Excel spreadsheets and the duplicated were eliminated. The available literature found was selected, inclusion and exclusion criteria were applied. The full articles selected were exported and the quality criteria were applied. Based on the full content of each selected article, the data extraction was subject of a critical analysis to seek literature answers for the two questions of this paper.

Table 2
**Database and Search Strings**

| Search ID | Scientific database | Search String |
|---|---|---|
| A | Web of Science | SEARCHED IN TOPICS<br><br>The review considers the following meanings: ("Central Bank Digital Currency" OR "CBDC") AND ("Information Technology") AND ("Governance"). |
| B | Scopus | SEARCHED IN TITLE, ABSTRACT AND KEYWORDS<br><br>The review considers the following meanings: ("Central Bank Digital Currency" OR "CBDC") AND ("Information Technology") AND ("Governance"). |
| C | Scopus | SEARCHED IN TITLE, ABSTRACT AND KEYWORDS<br><br>The review considers the following meanings: ("Central Bank Digital Currency" OR "CBDC") AND ("Information Technology" ). |

Source: Author.

III. Reporting the review

This item is written in the item four of this paper, Results and Discussion.

## 4. Results and Discussion

For Scopus database search it was necessary to apply a second search "C", this one excluding the string "Governance". In search "B" when using the term "governance" it was only found one study; *Surveillance technologies: Trends and social implications (Book Chapter) Lyon, D. 2004 The Security Economy 9789264107748, pp. 127-148.* This source was not considered once this does not reach the inclusion criteria of this paper. Table 3 shows the numbers.

Table 3
**Numbers found**

| Numbers Found | Web Of Science | Scopus¹ |
|---|---|---|
| Total Itens found | 110 | 11 |
| Available (Open access) | 48 | 7 |
| Considered for reading² | 10 | 2 |

Source: Author.

¹**Search C.**  ²**Reached the including criteria.**

Fourteen papers were fully considered for further analysis to answer the main questions of this paper. Twelve were selected through Literature Review following the including criteria, two additional papers were selected from the Bank for International Settlements (BIS) publications. It is important to mention that all of them are relatively new publication. The graphic 1 below points out the amount of the selected papers according to the respective year of publication, clearing showing that they are very new.

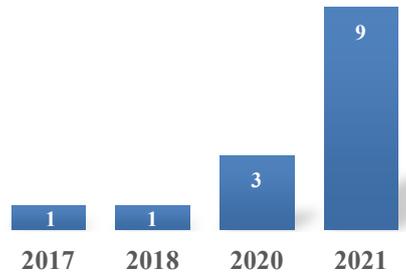

**Graphic 1.** Number of chosen papers and respective year of publication
Source: Author.

Table 4 below points out every paper considered in this study to answer the two proposed questions.

Table 4
**Literature selected to answer the two questions**

| Work Title | Source | Authors | Year | Source |
|---|---|---|---|---|
| Design Principles and Best Practices of Central Bank Digital Currency | Scopus | Dongcheng Li, W. Eric Wong, Sean Pan, Liang Seng Koh, and Matthew Chau. | 2021 | International Journal of Performability Engineering |
| Multi-Blockchain Model For Central Bank Digital Currency | Scopus | He Sun, Hongliang MAO, Xiaomin Bai, Zhidong Chen, Kai Hu, and Wei Yu. | 2017 | 18th International Conference on Parallel and Distributed Computing, Applications and Technologies (PDCAT) |
| Blockchains and distributed ledgers in retrospective and perspective | Web of Science | Alexander Lipton | 2018 | The Journal of Risk Finance |
| A Scientometric Review of Digital Currency and Electronic Payment Research: A Network Perspective | Web of Science | Qing Shi and Xiaoqi Sun | 2020 | Hindawi Complexity |
| A Compendium of Practices for Central Bank Digital Currencies for Multinational Financial Infrastructures | Web of Science | Edwin Ayisi Opare And Kwangjo Kim | 2020 | IEEE Access |
| A Hybrid Model for Central Bank Digital Currency Based on Blockchain | Web of Science | JINNAN ZHANG, RUI TIAN, YANGHUA CAO, XUEGUANG YUAN, ZEFENG YU, XIN YAN, AND XIA ZHANG. | 2021 | IEEE Access |
| Blockchain Application for Central Banks: A Systematic Mapping Study | Web of Science | NATALIA DASHKEVICH, STEVE COUNSELL, AND GIUSEPPE DESTEFANIS | 2020 | IEEE Access |
| Blockchain Implementation Method for Interoperability between CBDCs | Web of Science | Hyunjun Jung and Dongwon Jeong | 2021 | Future Internet |
| A global perspective on central bank digital currency | Web of Science | David Kuo Chuen Lee, Li Yan & Yu Wang | 2021 | China Economic Journal |
| New application of blockchain: digital currency model design in the context of Winter Olympics | Web of Science | Zihuan Feng and Xun Liang. | 2021 | MATEC Web of Conferences |
| Digitization and the evolution of money as a social technology of account | Web of Science | Michael Peneder | 2021 | Journal of Evolutionary Economics |
| Central Banks' Monetary Policy in the | Web of Science | Miguel Ángel Echarte Fernández, Sergio Luis | 2021 | Sustainability |

| Face of the COVID-19 Economic Crisis: Monetary Stimulus and the Emergence of CBDCs | | Náñez Alonso, Javier Jorge-Vázquez and Ricardo Francisco Reier Forradellas | | |
|---|---|---|---|---|
| BIS Working Papers No 948 Central bank digital currency: the quest for minimally invasive technology | Bank for International Settlements (BIS) | Raphael Auer and Rainer Böhme – Bank for International Settlements | 2021 | Bank for International Settlements – BIS – Working Paper. |
| BIS Working Papers No 941 The digitalization of money | Bank for International Settlements (BIS) | Markus K Brunnermeier, Harold James and Jean-Pierre Landau – Bank for International Settlements. | 2021 | Bank for International Settlements – BIS – Working Paper. |

Source: Author.

**Research Question 1- What IT resources and tools have been considered or applied to set the governance of CBDC adoption?**

After analyzing the fourteen papers, table 5 was organized to raise possible IT elements (these should be understood as IT resources, tools, elements, enablers, and any other possible IT item of utility in the searched field). Every IT resource, tool, functionality and feature can be considered as an IT element or enabler to a CBDC adoption, taking this into consideration the elements were raised through literature review.

Table 5
**Literature findings addressing the first research question**

| Elements | Authors |
|---|---|
| • Distributed Ledger Technology – DLT | Lia *et al* (2021); Opare and Kim, (2020). |
| • Smart Contracts and Programmability Design | Lia *et al* (2021). |
| • Single-tier and multi-tier operating model | Lia *et al* (2021). |
| • Cross-Blockchain Communication | Sun *et al* (2017). |
| • Multi-Blockchain Communication | Sun *et al* (2017). |
| • Blockchain and Distributed Ledgers | Lipton, (2018); Shi and Sun, (2020); Dashkevich *et al*, (2020). |
| • Blockchain | Opare and Kim, (2020); Fernandez *et al*, (2021). |
| • Assets Tokenization | Opare and Kim, (2020). |
| • Single Ledger and Cross Ledger | Opare and Kim, (2020). |
| • CBDC management method based on ISO/IEC 11179. The ISO/IEC 11179 metadata registry is an international standard for data interoperability. | Jung and Jeong, (2021). |
| • Blockchain and Interoperability | Jung and Jeong, (2021); Lee *et al*, (2021); Brunnermeier *et al*, (2021). |
| • Digital wallets | Auer and Böhme, (2021) |
| • Digital wallet infrastructure based on Blockchain | Feng and Liang, (2021). |
| • Digital payment instruments associated with platforms | Brunnermeier *et al*, (2021). |

|   | • Platform based economy and convertibility | Brunnermeier *et al*, (2021). |

Source: Authors.

Based on the experiences of China about CBDC, Lee *et al* (2021) list ten enabling factors. These factors are divided into three main groups. Table 6 below lists the enablers according to their respective group.

Table 6
**Enablers for CBDC mass adoption (Lee *et al*, 2021)**

| Factors of Optimal Integrated Ecosystem | Critical Enablers for mass adoption |
|---|---|
| Integrated and Enabling Infrastructure | 1-Digital Identification 2-Data Privacy Protection 3-Interoperable Value Transfer Gateway 4-Talent, Knowledge, and Skills |
| Global Cooperative Standard and Compliance | 5-Compliance Easy 6-Comprehensive Data and Oracle Ecosystem 7-Open-source and Trust Distribution Governance |
| Accessibility, Inclusivity of Storage, and Exchange | 8-Digital Literacy and User Experience 9-Strong Security Frameworks 10-Fast and Stable Network |

Source: Adapted from Lee *at al* (2021).

Interoperability is one of the most critical barriers to the adoption and function of CBDC, additionally it will allow multiple CBDCs to interact and exchange data with each other (Lee *et al*, 2021). Inclusivity and decentralization should be considered, according to Lee *et al* (2021) centralized governance and decentralized operation are not contradictory. They mention the 6Ds attributes, they are: digitization, disintermediation, decentralization, democratization, data privacy and disappearance (Lee *et al*, 2021).

Question 1 – General Discussion:

Although the theme of CBDC is indeed very new, the Literature already covers many It elements that are related to the use (and adoption) of CBDC and other digital currencies. IT elements can be understood as IT resources, tools, elements, enablers, and any other possible IT item of utility that allow the successful use of CBDC. Table five above groups many IT elements that according to the literature are helpful in the use of CBDC and other digital assets. Some authors mention Blockchain technology, others blockchain and other Distributed Ledger Technologies (DLTs). There are many studies that use some use cases and practical applications in order to better understand the IT resources needed in the adoption of CBDC.

Taking into consideration that there might exist many CBDC in the world, some authors mention the importance of some technology features, for instance cross-blockchain communication and multi-blockchain communication (Sun *et al*, 2017). Other important feature is the blockchain interoperability, which is closely related to the communication of the CBDC system as a whole, this feature is mentioned by Jung and Jeong, (2021); Lee *et al*, (2021); and Brunnermeier *et al*, (2021). Jung and Jeong (2021) considers the ISO/IEC 11179 metadata registry (an international standard for data interoperability) to consider in the DLT systems. Brunnermeie*r et al* (2021) mentions that the digital economy is a Platform based economy and convertibility should be considered (Brunnermeier *et al*, 2021).

Through this literature review it is possible to infer that not only the elements necessary for the CBDC adoption (like digital wallets and DLTs) but also the soft or functional operational skills (like convertibility and cross-communication among DLTs) have been raised by literature, what shows that we are in a more mature aspect about the IT elements needed for the successful CBDC use. It is important to mention that Lee *at al* (2021) listed ten critical enablers for CBDC mass adoption, among them there are compliance, security and governance enablers, table six details the authors consideration.

**Research Question 2- Identify IT governance models in the financial market due to CBDC adoption.**

Table 7 below points out the findings through the Literature Review that are potential answers for the second research question.

Table 7
**Literature findings addressing the second research question**

| Authors | Helpful findings to answer the question |
|---|---|
| Lia *et al* (2021) | Mention that three architectures have been identified: indirect CBDC, direct CBDC, and hybrid, a fusion of the previously mentioned two architectures. |
| Zhang *et al* (2021) | Proposed a network architecture based on a modular blockchain architecture, additionally a sliced data storage solution was designed to enhance the concurrency of this structured network. According to the authors the DPOS-BFT algorithm is optimized, which reduces the two rounds of consensus of the original algorithm to one round (Zhang *et al*, 2021). According to Zhang *et al* (2021) there are two major digital currency expression schemes: UTXO and Account. UTXO is an encrypted string with face value which seems as cash, and the scheme is used in Bitcoin (Zhang *et al*, 2021).<br><br>The authors proposed a hybrid blockchain system for CBDC, which is innovative in three levels: technology scheme, network architecture, and consensus mechanism. This hybrid model of UTXO and account improves the processing rate by 16.4% compared with UTXO. And the transaction processing speed is improved by 26.3% using the network architecture. With the improvement of the consensus algorithm, the consensus speed is improved by more than 51.8%. The authors pointed out that there are still some problems to be solved (Zhang *et al*, 2021). |
| Peneder (2021) | Mentions that the most ambitious project about CBDC seems to be the Digital Currency Electronic Payment system (CD/EP) initiated by the People's Bank of China (PBC). Digitizing the monetary base through a renminbi-backed form of electronic payment, according to him it will likely be the world's first comprehensive CBDC to become operational. The biggest goal is to reinforce the PBC's monetary authority and prevent the proliferation of other international digital currencies, while preserving intermediation within the traditional binary system of commercial and central banking (Peneder, 2021). |
| Lee *et al* (2021) | Lee *et al* (2021) proposes a two-tier or multi-tier ledger design and ten enablers of mass adoption and successful implementation of CBDC. This design allows central banks to manage the process flow, focus on the monitoring and control (Lee *et al*, 2021). The authors mention three key considerations for central banks in CBDC design processes. First, CBDC is guaranteed by the government and maintains its legal currency status. The government must directly support CBDC to ensure its tokenization. Second, the use and deposit of CBDC must not bear the corporate and credit risk of the custody entity or financial institution. Third, the cooperation between public and private sectors is necessary for the design of new CBDC (Lee *et al*, 2021). |
| Brunnermeier *et al*, BIS (2021) | Brunnermeier *et al* (2021) considers that in the digital economy interactions will occur within the borders of a "Digital Currency Area" (DCA), these areas will form endogenously and may or may not be governed by national boundaries (Brunnermeier *et al*, 2021). The |
| | authors define a digital currency area as a network where payments and transactions are made digitally by using a currency that is specific to that network. In a digital economy a regime in which all money is convertible to CBDC would uphold the unit of account status of public Money (Brunnermeier *et al*, 2021). |
| (Auer and Böhme – BIS, 2021) | Auer and Böhme (2021) state that every form of digital money requires a distributed record-keeping system. To them distributed con be understood as being implemented in many different places. Technical architecture of a CBDC is defined by the role of the components of the distributed record-keeping system, their communication relations, and who is in control of each component. Auer and Böhme (2021) list four possible architectures for CBDCs, they are: I) Direct CBDC, II) Hybrid CBDC, III) Intermediated CBDC, and IV) Indirect Architecture (Fully backed payment accounts or stablecoins) (Auer and Böhme, 2021). <br><br> The authors explain that in the "Direct CBDC", the CBDC is a direct claim on the central bank, which also handles all payments in real time keeping a record of all retail holdings. Hybrid CBDC architectures incorporate a two-tier structure with direct claims on the central bank while real-time payments are handled by intermediaries. Several variants of the hybrid architecture can be envisioned. The central bank could either retain a copy of all retail CBDC holdings or run a wholesale ledger. In the indirect architecture a CBDC is issued and redeemed only by the central bank, but this is done indirectly to intermediaries. The central bank operates the wholesale payment system only. About technological resilience, it is achieved in the hybrid design via the central bank operating a backup infrastructure. <br><br> (Auer and Böhme, 2021). Payment systems feature intermediaries but seek to reduce dependence on them. For example, many CBDC prototypes are based on enterprise versions of distributed ledgers, such as Corda, Hyperledger, or Quorum. Some academic proposals are designed to break with conventional databases. They adapt selected principles of decentralized cryptocurrencies to the use case of CBDC (Auer and Böhme, 2021). |

Source: Author.

Question 2 – General Discussion:

The main objective of this research question was to identify possible new IT governance models in the financial markets due to CBDC adoption and use, however according to the main findings it is difficult, or still embryonic to say that there is a definition in the literature (and in the business market) treating either the new IT infrastructure or IT governance model as a whole. It was possible to find some architecture patterns, operational mechanisms and design features. It was not found enough literature information in order to compare to a COBIT framework, like explained by Alreemy *et al* (2016) and Webb *et al* (2006).

Lia *et al* (2021), identified three architectures: indirect CBDC, direct CBDC, and hybrid (Lia *et al*, 2021). Zhang *et al* (2021) mentions some network architecture based on a modular Blockchain architecture. Additionally, there are some digital currency expression schemes: UTXO and Account and hybrid blockchain system (Zhang *et al*, 2021). Lee *et al* mention the two-tier or multi-tier ledger design (Lee *et al*, 2021). An important concept that may influence new governance models is stated by Brunnermeier *et al* (2021). They mention the "Digital Currency Area" (DCA) which considers that interactions in the digital economy will occur within the borders of them (DCAs). This may be a new governance feature that might take Institutions and geographical aspects to functions or positions not influencing the transactions.

A very promising paper is presented by the Bank for International Settlements (BIS), Auer and Böhme (2021). In the study the authors list and detail four possible architectures for CBDCs, they are: I) Direct CBDC, II) Hybrid CBDC, III) Intermediated CBDC, and IV) Indirect Architecture (Fully backed payment accounts or stablecoins) (Auer and Böhme, 2021).

According to this literature review it is difficult to say that there will be a ubiquitous IT governance model due to CBDC adoption, yet there are indeed new governance and/or architectural and/or operational features that are coming along with CBDC reality. It is most probably that new governance models are associated with use cases and practical evidence. This review shows that there is a need in the literature to keep fostering the research in this field, there is much yet to learn about CBDC associated with its IT governance.

In the Hybrid architecture interoperability would certainly be an important IT enabler feature. Although the literature in this field (CBDC and IT governance) is naturally embryonic, some IT elements and features have already been elucidated by researchers. For instance, interoperability, cross-communication, DLTs architecture and platform-based economy along with convertibility are all considered by literature as important features. These may indeed influence new governances.

## 5. Conclusions, Limitations and Future Works.

This paper has as main objective to answer two questions through a literature review. The first "*What IT resources and tools have been considered or applied to set the governance of CBDC adoption?*" and the second "*Identify new IT governance models in the financial market due to CBDC adoption*". It is possible to observe that this literature review points out important evidence. Although the theme of digital currencies, specially CBDC, is relatively new to the Literature, this paper allows to realize that the IT elements (which can be understood as IT resources, tools, among others) are already identifiable to literature. IT elements and features have already been elucidated by researchers. For instance, interoperability, cross-communication, DLTs architecture and platform-based economy along with convertibility are all considered by literature as important features. These may indeed influence new governances.

According to this literature review it is difficult to say (for the time being) that there will be a ubiquitous IT governance model due to CBDC adoption, yet there are indeed new governance and/or architectural and/or operational features that are coming along with CBDC reality. It is most probably that new governance models will be associated with use cases and practical evidence. This review shows that there is a need in the literature to keep fostering the research in this field, there is much yet to learn about CBDC associated with its IT governance.

It is possible to conclude that an IT governance model for CBDC may vary according to many variables, moreover both in the academic and in the business word, this is an embryonic theme which must be explored further. In the sum, it is possible to conclude that on one hand IT elements that may influence CBDC governance are in a more mature curve of knowledge by literature, on the other hand an IT governance model for CBDC is still growing and must be explored in many research aspects.

The fact that the theme of CBDC is very new to the literature could not be considered as a limiting factor for this study, however one of the including criteria was to find studies that cover together CBDC and Information Technology, and this could be seen as a limiting factor once the theme is new and there is a reduced number of available publications on databases. Future studies could keep exploring the theme of CBDC IT governance and their use cases. It is also helpful if future works study deeper the functions that each IT resource (or IT governance enabler for CBDC implementation) has in the CBDC IT architecture.

This paper contributes to the literature in the field once it brings possible IT elements (already recognized by literature) that are employed in the CBDC adoption. Secondly, this paper seeks to understand the IT governance of CBDC reuniting the main and the most updated articles in the field. It was found that the first question is more explored and in a more maturity level of comprehension by academics, however the second question deserves more attention and exploitation by the literature.

A research agenda should consider the following steps toward a deeper understanding; i) identify the role that each IT element has in the mechanisms of CBDC IT governance aspects; ii) explore both technical and managerial aspects about IT infrastructure mechanisms related to the CBDCs; and iii) describe the IT governance about the CBDCs (in preliminary use cases) considering their initial IT architectural aspects.

**REFERENCES**

A. Tapscott, and D. Tapscott, "How blockchain is changing finance," *Harvard Business Review*, vol. 1, no. 9, pp. 2-5, 2017.

Alreemy, Z., Chang, V., Walters, R., Wills, G. *Critical success factors (CSFs) for information technology governance (ITG)*. International Journal of Information Management 36. 907–916. 2016.

Amess, K.; Brown, S.; Thompson, S. *Management buy outs, supervision and employee discretion*. Scottish Journal of Political Economy, Hoboken, v. 54, n.4, p. 447-474, 2007.

Annual Economic Report BIS 2020.

Auer, R, G Cornelli and J Frost (2020): "Rise of the central bank digital currencies: drivers, approaches and technologies", *CEPR Discussion Paper* 15363, October, 2020.

Auer, R. and Böhme, R. *BIS – Bank for International Settlements. Working Papers No 948*. Central bank digital currency: the quest for minimally invasive technology. Monetary and Economic Department. June 2021.

Bank for International Settlements (BIS) (2020), "Central banks and payments in the digital era", *BIS Annual Economic Report*, Ch. III, June.

Borio, C. (2019) "On Money, Debt, Trust and Central Banking"; BIS Working Paper No. 763; February.

Carstens, A (2020): "Shaping the future of payments", *BIS Quarterly Review*, March, pp. 17–20.

Chaum, D (1985): "Security without identification: transaction systems to make big brother obsolete", *Communications of the ACM*, vol 28, no 10, pp 1030–1044.

Chi, M., Zhao, J., George, J. F., Li, Y., Zhai, S. *The influence of inter-firm IT governance strategies on relational performance: The moderation effect of information technology ambidexterity*. International Journal of Information Management 37. 43–53. 2017.

D. Dalal, S. Yong, and A. Lewis, "The Future is Here–Project Ubin: SGD on Distributed Ledger," *Monetary Authority of Singapore & Deloitte*, 2017.

David Kuo Chuen Lee, Li Yan & Yu Wang (2021) A global perspective on central bank digital currency, *China Economic Journal*, 14:1, 52-66, DOI: 10.1080/17538963.2020.1870279. 2021.

Dongcheng Lia , W. Eric Wonga, Sean Panb , Liang Seng Kohb , and Matthew Chaua. Design Principles and Best Practices of Central Bank Digital Currency. *International Journal of Performability Engineering.* Available online at www.ijpe-online.com vol. 17, no. 5, pp. 411-421. 2021.

E. Androulaki *et al*., "Hyperledger fabric: a distributed operating system for permissioned blockchains," in Proc. *13th EuroSys Conference*, 2018, pp. 1-15.

Echarte Fernández, M.Á.; Náñez Alonso, S.L.; Jorge-Vázquez, J.; Reier Forradellas, R.F. Central Banks' Monetary Policy in the Face of the COVID-19 Economic Crisis: Monetary Stimulus and the Emergence of CBDCs. *Sustainability* 2021, 13, 4242.

European Central Bank, Bank of Japan. Securities settlement systems: delivery-versus-payment in a distributed ledger environment. [Online]. Available: https://www.ecb.europa.eu/pub/pdf/other/stella_project_report_marc h_2018.pdf

G. Danezis, and S. Meiklejohn, "Centrally banked cryptocurrencies," arXiv preprint arXiv:1505.06895, 2015.

Gilbert, S and N Lynch (2002): "Brewer's conjecture and the feasibility of consistent, available, partitiontolerant web services", ACM SIGACT News, vol 33, no 2, pp 51-59.

Haber, S.; Stornetta, W.S. How to time-stamp a digital document. In Conference on the Theory and Application of Cryptography; Springer: Berlin, Germany, 1990; pp. 437–455.
He Sun Hongliang MAO, Xiaomin Bai, Zhidong Chen, Kai Hu, Wei Yu. Multi-Blockchain Model for Central Bank Digital Currency.*18th International Conference on Parallel and Distributed Computing, Applications and Technologies (PDCAT).* Beijing, 100191, China, 2017.

J. Chapman, R. Garratt, S. Hendry, A. McCormack, and W. McMahon, "Project Jasper: Are distributed wholesale payment systems feasible yet," *Financial System*, vol. 59, 2017.

JINNAN ZHANG, RUI TIAN , YANGHUA CAO, XUEGUANG YUAN, ZEFENG YU, XIN YAN, AND XIA ZHANG.A Hybrid Model for Central Bank Digital Currency Based on Blockchain. *IEEE Access*, v9, 2021.

Jung, H.; Jeong, D. Blockchain Implementation Method for Interoperability between CBDCs. *Future Internet* 2021, 13, 133. https://doi.org/10.3390/fi13050133. 2021.

Ju Wu, S. P., Straub, D. W., Liang, T. P. *How Information Technology governance mechanisms and strategic alignment influence organizational performance: Insights from a matched survey of business and IT managers*. MIS Quarterly Vol. 39 No. 2, pp. 497-518, 2015.

Kiff, J., J. Alwazir, S. Davidovic, A. Farias, A. Khan, T. Khiaonarong, M. Malaika, *et al*. 2020. A Survey of Research on Retail Central Bank Digital Currency. Washington, DC: *International Monetary Fund* (IMF).

Kitchenham, B. (2004). Procedures for performing systematic reviews. Keele, UK, Keele University, 33(2004), 1–26.

Kitchenham, B., Brereton, O. P., Budgen, D., Turner, M., Bailey, J., & Linkman, S. (2009). Systematic literature reviews in software engineering: A systematic literature review. *Information and Software Technology*, 51(1), 7–15.

L. Sun, "Central bank digital currencies," *Journal of Digital Banking*, vol. 4, no. 1, pp. 85-94, 2019.

Lipton, A. Blockchains and distributed ledgers in retrospective and perspective. *The Journal of Risk Finance* Vol. 19 No. 1,pp. 4-25, 2018.

M. L. Bech, and R. Garratt, "Central bank cryptocurrencies," *BIS Quarterly Review* September 2017.

Markus K Brunnermeier, Harold James and Jean-Pierre Landau. The digitalization of money. *BIS – Bank for International Settlements. Working Papers No 941. Monetary and Economic Department*, May 2021.

NATALIA DASHKEVICH, STEVE COUNSELL, AND GIUSEPPE DESTEFANIS. Blockchain Application for Central Banks: A Systematic Mapping Study. *IEEE Access*, 2020.

Opare, E. A., and Kim, K.A. Compendium of Practices for Central Bank Digital Currencies for Multinational Financial Infrastructures. *IEEE Access*, 2020.

Peneder, M. Digitization and the evolution of money as a social technology of account. *Journal of Evolutionary Economics*, 2021.

Qing Shi and Xiaoqi Sun. Review Article A Scientometric Review of Digital Currency and Electronic Payment Research: A Network Perspective. *Hindawi Complexity*.https://doi.org/10.1155/2020/8876017.2020.

Sveriges Riksbank (2019): "The Riksbank proposes a review of the concept of legal tender", press announcement.

Tranfield, D., Denyer, D., & Smart, P. (2003). Towards a methodology for developing evidence-informed management knowledge by means of systematic review. *British Journal of Management*, 14(3), 207–222.

Turel, O., Liua, P., Bartb, C. *Board-Level Information Technology Governance Effects on Organizational Performance: The Roles of Strategic Alignment and Authoritarian Governance Style*. Information Systems Management. Vol. 34, N. 2, 117–136, 2017.

Webb, P.; Pollard, C.; Ridley, G. *Attempting to define IT governance: wisdom or folly?* In: Hawaii International Conference on System Sciences, 39., Kauai, 2006. Proceedings. Kauai: Shidler College of Business, 2006. p. 10.

W.-T. Tsai, Z. Zhao, C. Zhang, L. Yu, and E. Deng, "A Multi-Chain Model for CBDC," in Proc. *5th International Conference on Dependable Systems and Their Applications* (DSA), Sep. 2018, pp. 25- 34.

Zihuan Feng, and Xun Liang. New application of blockchain: digital currency model design in the context of Winter Olympics. *MATEC Web of Conferences* 336, 08011 (2021).